\documentclass{elsart}

\usepackage{amssymb,epsfig,subfigure}

\def\ifm#1{\relax\ifmmode#1\else$#1$\fi}

\def\x{\ifm{\times}}

\let\cal=\mathcal   
\def\ORD#1!{\ifm{{\cal O}\hbox{(#1)}}}
\def\pt#1,#2,{\ifm{#1\x10^{#2}}} 

\def\bye{\end{document}}

\newcount\figurecount    
\def\be{\begin{equation}}
\def\ee{\end{equation}}  
\def\bea{\begin{eqnarray}}
\def\eea{\end{eqnarray}}
\def\bc{\begin{center}}
\def\ec{\end{center}}  
\def\ben{\begin{enumerate}}
\def\een{\end{enumerate}}

\def\sms{\kern-1mm}

\makeatletter
\newdimen\z@ \z@=0pt 
\newskip\z@skip \z@skip=0pt plus0pt minus0pt
\def\m@th{\mathsurround=\z@}
\def\ialign{\everycr{}\tabskip\z@skip\halign} 
\def\eqalign#1{\null\,\vcenter{\openup\jot\m@th
  \ialign{\strut\hfil$\displaystyle{##}$&$\displaystyle{{}##}$\hfil
      \crcr#1\crcr}}\,}
\makeatother

\newcommand{\aff}[2]{Dipartimento di Fisica dell'Universit\`a #1 e Sezione INFN, #2, Italy.}
\newcommand{\affd}[1]{Dipartimento di Fisica dell'Universit\`a e Sezione INFN, #1, Italy.}

\begin{document}

\begin{frontmatter}


\title{Measurement of the ratio 
$\Gamma(K_L \to \gamma \gamma)  / {\Gamma(K_{L} \to \pi^{0} \pi^{0}
  \pi^{0}) }$ 
 with the KLOE detector}

\vskip -1cm

\collab{The KLOE Collaboration}

\author[Roma2]{M.~Adinolfi},
\author[Na] {A.~Aloisio},
\author[Na]{F.~Ambrosino},
\author[Frascati]{A.~Antonelli},
\author[Frascati]{M.~Antonelli},
\author[Roma3]{C.~Bacci},
\author[Frascati]{G.~Bencivenni},
\author[Frascati]{S.~Bertolucci},
\author[Roma1]{C.~Bini}
\author[Frascati]{C.~Bloise}
% ...

\clearpage
\address[Bari]{\affd{Bari}}
\address[Frascati]{Laboratori Nazionali di Frascati dell'INFN, 
Frascati, Italy.}
\address[Karlsruhe]{Institut f\"ur Experimentelle Kernphysik, 
Universit\"at Karlsruhe, Germany.}
\address[Lecce]{\affd{Lecce}}
\address[Na]{Dipartimento di Scienze Fisiche dell'Universit\`a 
``Federico II'' e Sezione INFN,
Napoli, Italy}
\address[Pisa]{\affd{Pisa}}
\address[Energ]{Dipartimento di Energetica dell'Universit\`a 
``La Sapienza'', Roma, Italy.}
\address[Roma1]{\aff{``La Sapienza''}{Roma}}
\address[Roma2]{\aff{``Tor Vergata''}{Roma}}
\address[Roma3]{\aff{``Roma Tre''}{Roma}}
\address[StonyBrook]{Physics Department, State University of New 
York at Stony Brook, USA.}
\address[Trieste]{\affd{Trieste}}
\address[Virginia]{Physics Department, University of Virginia, USA.}
\address[Beijing]{Permanent address: Institute of High Energy 
Physics, CAS,  Beijing, China.}
\address[Moscow]{Permanent address: Institute for Theoretical 
and Experimental Physics, Moscow, Russia.}
\address[Tbilisi]{Permanent address: High Energy Physics Institute, Tbilisi
  State University, Tbilisi, Georgia.}

\begin{abstract}

\noindent 
We have measured the ratio 
$R = \Gamma(K_{L} \to \gamma \gamma)/ 
\Gamma(K_{L}\to \pi^{0} \pi^{0} \pi^{0})$ using the KLOE detector.
From a sample of $\sim 10^9$ $\phi$-mesons  produced
at DA$\Phi$NE, the Frascati $\phi-$factory, we select
$\sim 1.6 \ 10^{8}$  $K_L$-mesons tagged by observing $K_S \to \pi^+ \pi^-$
following the reaction $e^+ e^- \to \phi \to K_L K_S$. 
From this sample we select
27,375 $K_L \to \gamma \gamma$ events and obtain
$R   = (2.79 \pm 0.02_{stat} \pm 0.02_{syst}) \times 10^{-3}$.
Using the world average value for  $BR(K_{L} \to \pi^{0} \pi^{0} \pi^{0})$, we
obtain $BR(K_{L} \to \gamma \gamma) = (5.89 \pm 0.07 \pm 0.08) \times
10^{-4}$ where the
second error is due to the 
uncertainty on the $\pi^0 \pi^0 \pi^0 $ branching fraction.

\par\noindent PACS:
\par\noindent keywords:

\end{abstract}
\end{frontmatter}

\clearpage
\section{Introduction and experimental setup}

The decays $K_{S} \to \gamma \gamma$ and $K_{L} \to \gamma \gamma$ provide interesting
tests \cite{CHPT-1} of chiral perturbation theory, ChPT. The
dominant contribution 
to the $K_{S} \to \gamma \gamma$ decay is
$\mathcal{O}(p^{4})$ and can therefore be computed with 
reasonable accuracy in ChPT. 
The $\mathcal{O}(p^{4})$ term vanishes
for $K_{L} \to \gamma \gamma$ in the SU(3) limit. 
However large $\mathcal{O}(p^{6})$ contributions
mediated by pseudoscalar mesons \cite{CHPT-2} are expected for $K_L \to
\gamma \gamma$ with values 
depending on the amount of singlet-octet mixing \cite{CHPT-3}.
A precise measurement of the $K_{L} \to \gamma \gamma$ decay rate
is also of interest in connection with the $K_{L} \to \mu^{+}
\mu^{-}$ decay. In fact the absorptive part of the decay rate,
$\Gamma(K_{L} \to \mu^{+} \mu^{-})_{abs}$, is proportional to
$\Gamma(K_{L} \to \gamma \gamma)$.
This constrains the dispersive part, $\Gamma(K_{L} \to
\mu^{+} \mu^{-})_{dis}$ and eventually the possibility of determining
the V$_{td}$ parameter of the CKM matrix \cite{CHPT-1}. 
Measurements of $\Gamma(K_{S} \to \gamma
\gamma)/\Gamma(K_{S} \to \pi^{0} \pi^{0})$ and $\Gamma(K_{L} \to
\gamma \gamma)/\Gamma(K_{L} \to \pi^{0} \pi^{0} \pi^{0})$ have 
been recently published by the NA48 Collaboration
\cite{NA48-2002}. We describe a new measurement of $\Gamma(K_{L}
\to \gamma \gamma)/\Gamma(K_{L} \to \pi^{0} \pi^{0} \pi^{0})$ obtained
with $K_{L}$-mesons from $\phi \to K_S K_L$ decays at DA$\Phi$NE, the
Frascati $\phi-$factory.

In DA$\Phi$NE the electron and positron beams have energy $E =
m_{\phi}/(2 \cos\theta)$ where $\theta$ = 12.5 mrad is half of the 
beam crossing angle. 
$\phi$-mesons are produced with a cross section of
$\sim$ 3 $\mu$b and a momentum of 12.5 MeV/c toward the center of the
rings.

The center of mass energy, $W$,
the position of the beam crossing point ($x,y,z$) and the $\phi$
momentum are determined by measuring Bhabha
scattering events. In a typical run of integrated luminosity 
$\int {\cal L}$dt $\sim$ 100
nb$^{-1}$, lasting about 30 minutes,
we have $\delta W = 40$ keV, $\delta p_{\phi} = 30$
keV/c, $\delta x = 30~ \mu$m, and $\delta y = 30~ \mu$m.

The detector consists of a large cylindrical drift chamber, DC
\cite{K-DC}, surrounded by a lead-scintillating fiber sampling
calorimeter, EMC \cite{K-EMC},  both immersed in a solenoidal magnetic
 field of 0.52 T with the axis parallel to the beams.
The DC tracking volume extends from 28.5 to 190.5 cm
in radius and is 340 cm in length. For charged particles the
transverse momentum resolution is $\delta p_{T}/p_{T} \simeq
0.4\%$ and vertices are reconstructed with a spatial resolution of
$\sim$ 3 mm. 
The calorimeter is divided into a barrel and two endcaps
and covers 98$\%$ of the solid angle. 
Photon energies and
arrival times are measured with resolutions $\sigma_{E}/E =
0.057/{\small \sqrt{E \ ({\rm GeV})}}$  and $\sigma_{t} = 54 \ {\rm ps}
/{\small \sqrt{E \ ({\rm GeV})}} \oplus 50 \ {\rm ps}$  respectively. 
The photon entry points are determined with an accuracy 
 $\sigma_l \sim  1 \ {\rm cm} /{\small \sqrt{E \ ({\rm GeV})}}$ 
along the fibers, and $ \sim 1 $ cm in the transverse directions.
A photon is defined as a calorimeter cluster not associated to a charged 
particle, by requiring that the distance along the fibers
between the cluster centroid and the impact point of the nearest 
extrapolated track be greater than 3$\sigma_l$.
Two small calorimeters, QCAL \cite{K-QCAL}, made with lead and
scintillating tiles are wrapped around the low-beta quadrupoles to
complete the hermeticity.

The trigger \cite{K-TRIGGER} uses information from both the
calorimeter and the drift chamber. The EMC trigger requires
two local energy deposits above threshold 
($E > 50$ MeV in the barrel, $E > 150$ MeV in the endcaps). 
Recognition and rejection of cosmic-ray events is also performed at the
trigger level, checking for the presence of two energy deposits above 30
MeV in the outermost calorimeter plane.
The DC trigger is based on the multiplicity
and topology of the hits in the drift cells. 
The trigger has a large time spread  with respect to the beam crossing
time.
It is however synchronized with the machine radio frequency divided by four,
$T_{\rm sync}$ = 10.85 ns, with an accuracy of 50 ps.
During the period of data taking the bunch crossing period at 
DA$\Phi$NE was $T$ = 5.43 ns. 
The $T_0$ of the bunch crossing producing an event
is determined offline during the event reconstruction.

\section{Data analysis}
The $\phi$-meson decays into $ K_{S} K_{L}$
$\sim 34\%$ of the time. The production of a $K_{L}$ is tagged
by the observation of a $K_{S} \to \pi^{+} \pi^{-}$ decay.
$K_{L} \to \gamma \gamma$ and $K_{L} \to \pi^{0} \pi^{0} \pi^{0}$
decay vertices are reconstructed along the direction opposite 
to the $K_{S}$ in the $\phi$ rest frame 
and required to be inside a given fiducial volume, $FV$.
We call $R = N_{\gamma \gamma}/N_{\pi^{0}\pi^{0} \pi^{0}}$
the ratio of interest.
The numerators and denominators are found from:

\[ N = \frac{N_{obs} - N_{bgd}}
   {\epsilon_{trig} \cdot \epsilon_{tag} \cdot \epsilon_{FV}
   \cdot \epsilon_{sel}}
   \]
where $N_{obs}$ and $N_{bgd}$ are the numbers of observed events and estimated
background, $\epsilon_{trig}$, $\epsilon_{tag}$,
$\epsilon_{FV}$ and $\epsilon_{sel}$ are respectively the trigger
efficiency, the tagging efficiency, the acceptance in the fiducial
volume and the selection efficiency for the two decays. The
efficiencies $\epsilon_{tag}$ and $\epsilon_{trig}$
are equal at the few per mil level and cancel in the
ratio $R$. Background and selection efficiencies
must be separately determined.

For this analysis the drift chamber is used to measure the $K_{S}
\to \pi^{+} \pi^{-}$ decay and to determine the direction of the
$K_{L}$, the calorimeter is used to measure the photon energies
and impact points and to reconstruct the $K_{L}$ decay vertex 
by time of flight.

The data sample was collected during 2001 and 2002 for
an integrated luminosity of $\sim$ 362 pb$^{-1}$ corresponding to the
production of $\sim 10^{9}$ $\phi$. Details of the analysis can be
found in reference \cite{K-note}. $K_{L} \to \gamma \gamma$ events
have a very clear signature, being the only source of $\sim$ 250
MeV photon pairs that balance the momentum of the observed $K_{S}$.
This allows the use of very loose selection criteria.
On the other hand $K_{L} \to \pi^{0} \pi^{0} \pi^{0}$, the
dominant neutral decay, is characterised by a large multiplicity
of lower energy photons.
The final error on $R$ is dominated by the
error on the number of $K_{L} \to \gamma \gamma$ events.

Before full event reconstruction, the data are passed through a
filter to reject machine background and cosmic ray events.
As discussed later, this filter has a modest impact 
on the events of interest for this analysis.

$K_{S} \to \pi^{+} \pi^{-}$ decays are
selected with the following requirements:
\begin{itemize}
\item[-] two tracks with opposite charge that form a vertex 
with cylindrical coordinates 
$r_{v} < 4$ cm, $|z_v| < 8$ cm, and no other tracks connected
to the vertex;
\item[-] $K_S$ momentum 
100 MeV/c $ < \vec{p_{K_S}} = \vec{p}_{\pi^+} + \vec{p}_{\pi^-}< $ 120 MeV
in the $\phi$ rest system, and $\pi^{+} \pi^{-}$ 
invariant mass 490 MeV $< M_{\pi^{+} \pi^{-}} <$ 505 MeV.
\end{itemize}

The $K_S \to \pi^+ \pi^-$ decay 
provides an unbiased tag for the $K_{L}$ when it
decays into neutral particles and a good measurement of the
$K_{L}$ momentum, $\vec{p}_{K_{L}} = \vec{p}_{\phi} -
\vec{p}_{K_{S}}$, where $\vec{p}_{\phi}$ is the central value of the
$e^{+} e^{-}$ momentum determined with Bhabha scattering events. The angular
resolution on the $K_{L}$ direction is 
determined from $K_{L} \to \pi^{+} \pi^{-} \pi^{0}$ 
events by measuring the angle between $\hat{p}_{K_{L}}$ and the line 
joining the $\phi$ vertex and the $\pi^{+} \pi^{-}$ reconstructed vertex.
The widths of the angular distributions
are $\delta \phi = 1.5^{\circ}$, $\delta \theta = 1.8^{\circ}$.

\begin{figure}[htb]
    \centering
    \includegraphics[width = 7cm]{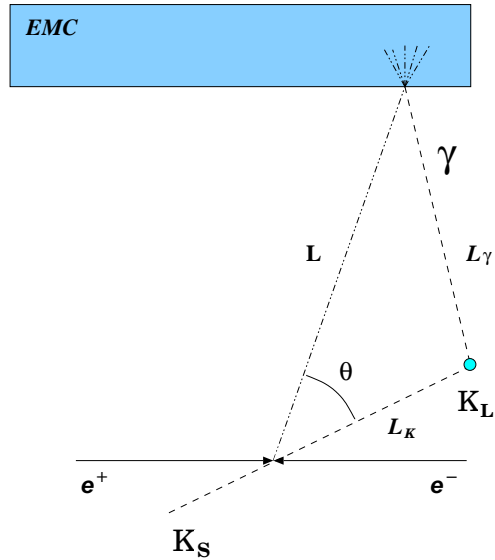}
    \caption{The time of flight triangle}
    \label{FIG1}
\end{figure}

The position of the $K_L$ vertex for $K_{L} \to \gamma \gamma$
and $K_{L} \to \pi^{0} \pi^{0} \pi^{0}$ decays is measured using the photon
arrival times on the EMC.
Each photon defines a time of flight
triangle shown in Fig. \ref{FIG1}. The three sides are the $K_{L}$
decay length, $L_{K}$; the distance from the decay vertex to the
calorimeter cluster centroid, $L_{\gamma}$; and the distance from
the cluster to the $\phi$ vertex, $L$. The equations to determine
the unknowns $L_K$ and $L_{\gamma}$ are:
\[ \begin{array}{rrr}
   L^2 + L_K^2 -2 L L_K \cos\theta & = & L_{\gamma}^2 \\
   L_K/\beta_K + L_{\gamma} & = & ct_{\gamma}
   \end{array} \]
where $t_{\gamma}$ is the photon arrival time on the EMC, $\beta_{K} c $ is the
$K_{L}$ velocity and $\theta$ is the angle between $\vec{L}$ and
$\vec{L}_{K}$. Only one of the two solutions is kinematically correct.
The value of $L_{K}$ is obtained from the energy
weighted average, $L_{K} = \sum_{i} E_{i} \cdot L_{Ki}/ \sum_{i}
E_{i}$  where {\it i} is the photon index. 

The accuracy of this method is checked by
comparing in the $K_{L} \to \pi^{+} \pi^{-} \pi^{0}$ decays
the position of the $K_{L}$ decay vertex measured both by timing with the
calorimeter and, with a much better
precision, by tracking with the drift chamber.

The resolution function, $\sigma(L_{K})$, 
is determined  with  $K_{L} \to \pi^{0} \pi^{0} \pi^{0}$ 
and $K_L \to \gamma \gamma$ events by the distribution of
the residuals $L_{K,i} - L_{K}$ where $L_{K}$ is the
average obtained with all the photons but the $i^{th}$.  
We measure for the $K_{L} \to \pi^{0} \pi^{0} \pi^{0}$  sample
$\sigma_{\pi^0 \pi^0 \pi^0}(L_{K}) =
2.06 - (0.16 \cdot 10^{-2} \ L_{K}) + ( 0.19 \cdot 10^{-4} \ L_{K}^2)$ (cm) 
and for the $K_{L} \to \gamma \gamma$ sample
$\sigma_{\gamma \gamma}(L_{K}) = 1.73 + 0.0033 \ L_{K}$ (cm).

The $K_{L}$ FV is defined in
cylindrical coordinates as 30 cm $< r_{t} <$ 170 cm, $|z| <$ 140 cm.
The fraction of $K_L$-mesons decaying in the FV is $(31.5 \pm 0.1) \%$.

The identification of the bunch crossing that originated the event
is crucial to locate the vertex in space.
An error by one bunch crossing period results in a displacement of
the $K_L$ vertex of about 33 cm and decreases the probability of
correctly associating the photon clusters.
The bunch crossing is determined by identifying one of the 
two pions of the $K_S$ decay and by measuring its track length, 
momentum and time of flight. 
Thus an
error of one (or more) crossing periods can occur if there is an
incorrect track-to-cluster association or the track parameters are
poorly measured.

To minimise the number of events with an incorrect bunch-crossing
assignment, we perform a consistency check of the time of flight
of the pions along their trajectory $l_{\pi}$ measured with the DC,
$t_{DC} = l_{\pi}/\beta_{\pi} \gamma_{\pi} c $ with
the corresponding cluster time measured by the
calorimeter, $t_{EMC}$. 
Requiring $|t_{DC} - t_{EMC}| < 2$ ns for at least one pion, 
the probability of correctly identifying
the bunch crossing is $(99.4 \pm 0.1) \%$. This
additional cut retains $96\%$ of the original $K_{S} \to \pi^{+}
\pi^{-}$ event sample.
The pro\-ba\-bi\-li\-ty of identifying the correct bunch crossing was measured
with a sample of $K_{L} \to \pi^{+} \pi^{-} \pi^{0}$ decays where
the position of the $\pi^+ \pi^-$ vertex, $r_{\pi \pi}$, is
reconstructed by tracking in the DC and the position of the two-photon
vertex, $r_{\gamma \gamma}$, by timing with the EMC. The
difference $r_{\pi \pi} - r_{\gamma\gamma}$ is used to isolate the
events in which the bunch crossing is incorrectly determined.

\section{$K_{L} \to \pi^{0} \pi^{0} \pi^{0}$ selection}

The $K_{L} \to \pi^{0} \pi^{0} \pi^{0}$ decay has a large
branching fraction, 21\%, and thus has very small background. Given
the large statistics we retain only 1 out of 10 decays.
The selection of $K_{L} \to \pi^{0} \pi^{0} \pi^{0}$ events requires
at least three calorimeter clusters with the following properties
\begin{itemize}
\item[-] energy larger than 20 MeV;
\item[-] distance from any other cluster larger than 40 cm;
\item[-] no association to a charged track;
\item[-] $L_{K}$ in the fiducial volume and
$|L_{Ki} - L_{K}| < 4 \sigma(L_{K})$.
\end{itemize}

The main sources of inefficiencies are: 1) geometrical acceptance;
2) cluster energy threshold; 3) merging of clusters; 4) accidental
association to a charged track; 5) Dalitz decay of one or more
$\pi^{0}$'s. The effect of these inefficiencies is to modify the
relative population for events with 3, 4, 5, 6, 7 and $\ge 8$,
clusters without significant loss of efficiency.
Monte Carlo simulation shows that the selection efficiency 
is $\epsilon_{sel} = (99.80 \pm 0.01)\%$.

A comparison between data and Monte Carlo of the relative
populations and of the distribution of the total energy, $E =
\sum_{i} E_{i}$, shows that only events with 3 and 4 clusters are
contaminated by background. This is due to 
$K_{L} \to \pi^{+} \pi^{-} \pi^{0}$ decays where one or two
charged pions produce a cluster not associated to a track and 
neither track is associated to the $K_L$ vertex or to $K_{L} \to
\pi^{0} \pi^{0}$ decays, possibly in coincidence with machine
background particles ($e^{\pm}$ or $\gamma$) that shower in the
QCAL and generate soft neutral particles.

To reduce this background, for the 3-cluster population 
we further require at least two
clusters in the barrel with at least one of them with energy $E >
50$ MeV and for the 4-cluster population at least one cluster in
the barrel with energy $E > 50$ MeV.
The probability to have a cluster with $E > 50 $ MeV has been
evaluated using the 6-cluster events. The probability of having a given
number of clusters in the barrel depends only on geometry and has
been evaluated by Monte Carlo simulation. We obtain
$\epsilon_{3, E>50  {\rm MeV} } = (81.9 \pm 0.1)\%$ and $\epsilon_{4,
  E>50{\rm MeV}} = (98.3 \pm 0.1)\%$. 
Additionally, an event with 3 clusters is accepted
only if an additional cluster is found in QCAL within a time
window of 10 ns with respect to the $K_{L}$ decay time. The probability of
such an occurrence is  $\epsilon_{qcal} = (52 \pm 2) \%$. 

The $K_{L} \to \pi^{+} \pi^{-} \pi^{0}$ background is
rejected by imposing a veto ({\em track veto}) on the events with
charged tracks not associated to the $K_{S}$ decay and with the
first hit in the drift chamber at a distance of less than 30 cm from the position of
the $K_{L}$ vertex. The track veto also rejects about $60\%$ of the
$K_{L} \to \pi^{0} \pi^{0} \pi^{0}$ events with Dalitz decays.

\begin{figure}[htb]
    \vfill\begin{minipage}{0.5\linewidth}
    \centering \epsfig{file=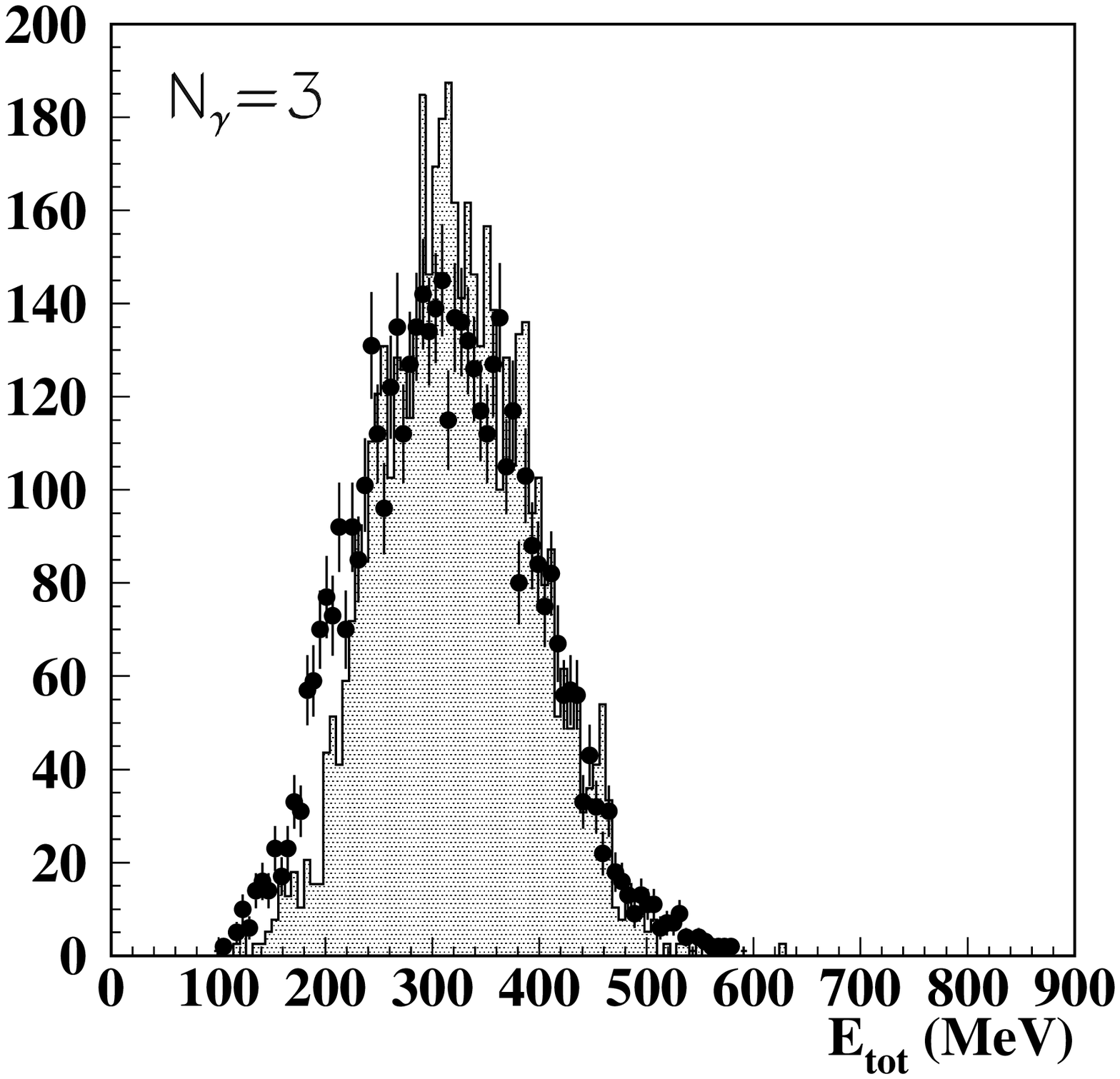,width=6cm}
\end{minipage}
\begin{minipage}{0.5\linewidth}
    \centering \epsfig{file=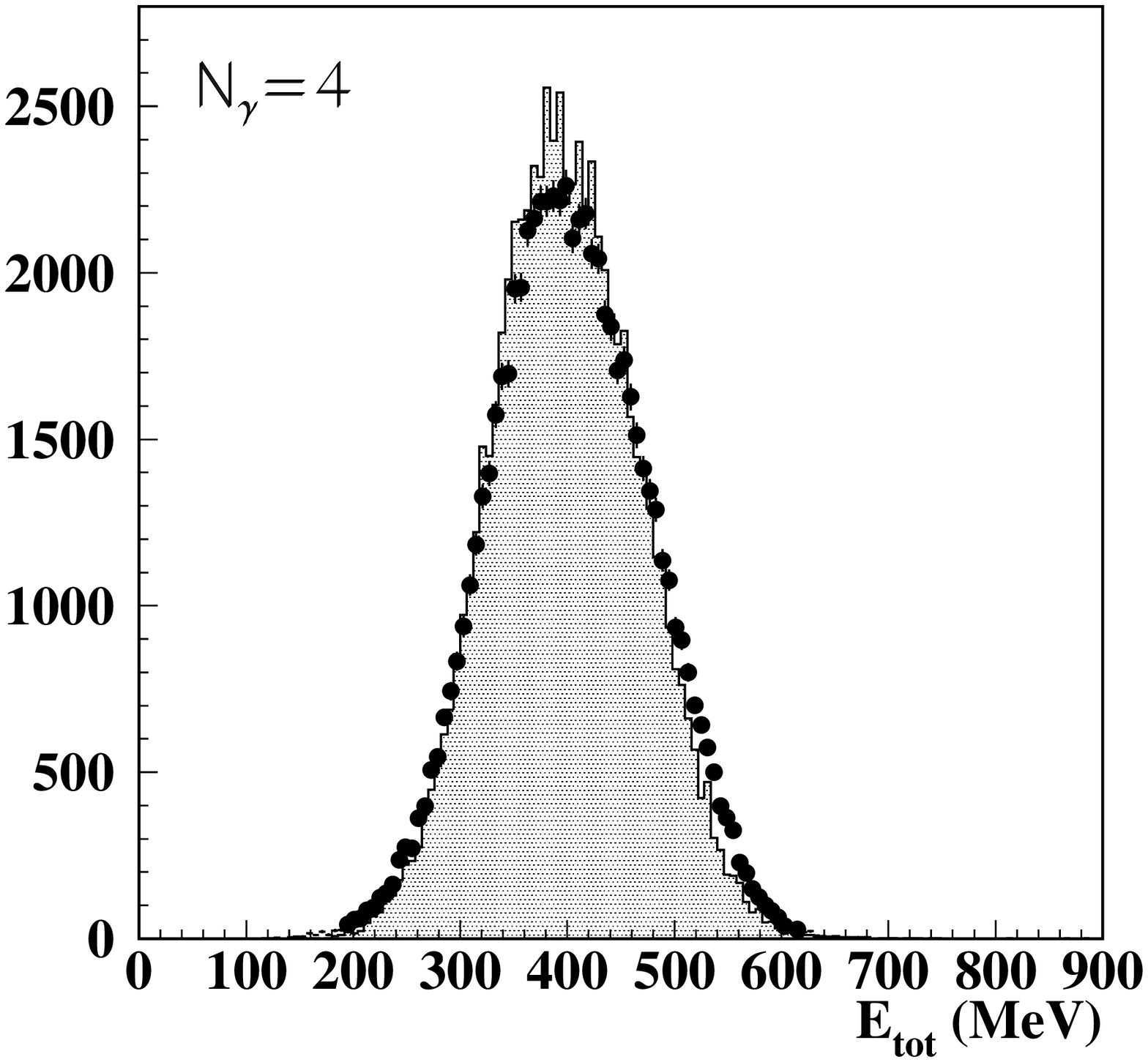,width=6cm}
\end{minipage}
    \vfill\begin{minipage}{0.5\linewidth}
    \centering  \epsfig{file=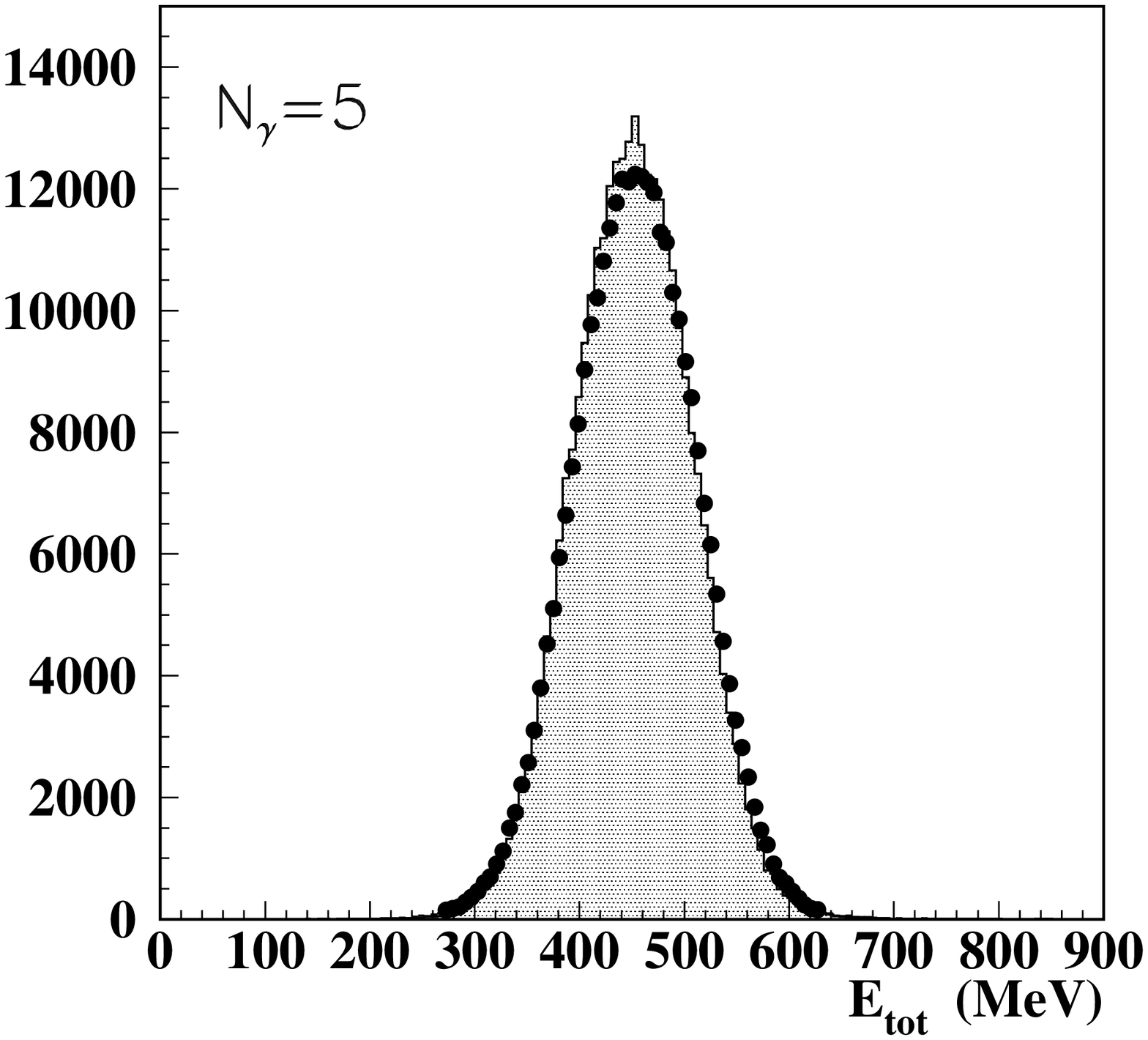,width=6cm}
\end{minipage}
\begin{minipage}{0.5\linewidth}
\centering  \epsfig{file=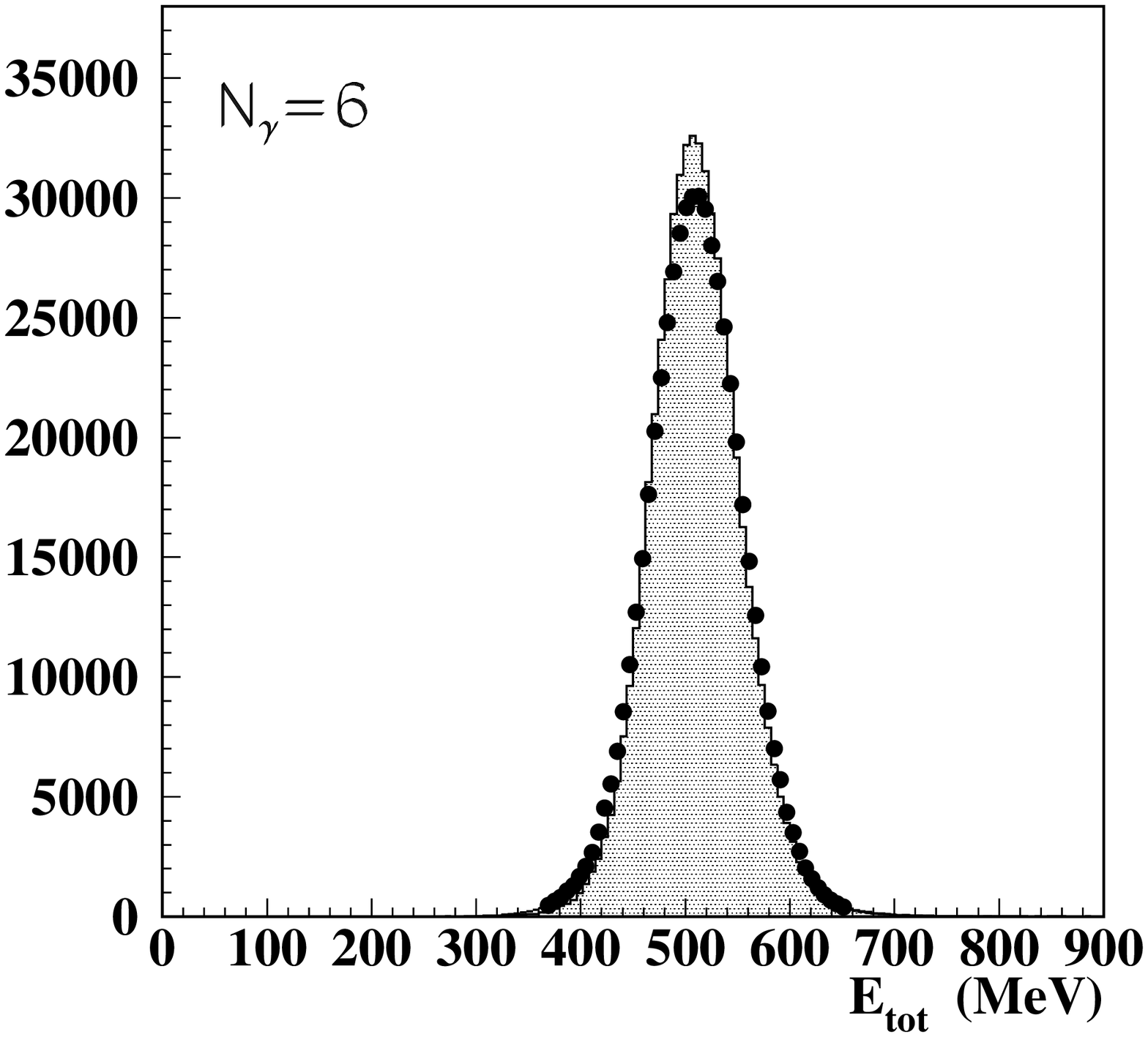,width=6cm}
\end{minipage}
 \vfill\begin{minipage}{0.5\linewidth}
   \centering  \epsfig{file=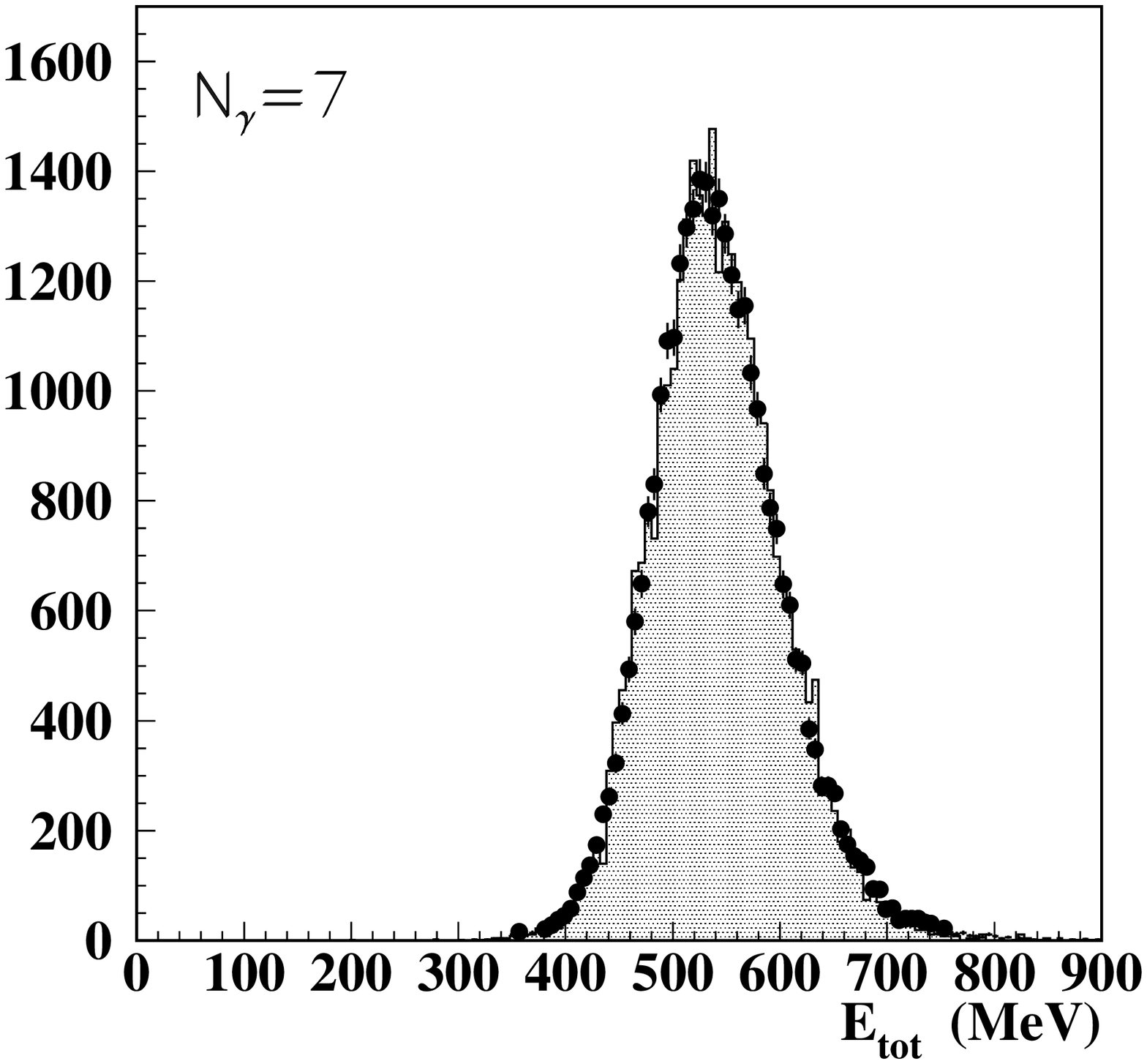,width=6cm}
\end{minipage}
\begin{minipage}{0.5\linewidth}
 \centering \epsfig{file=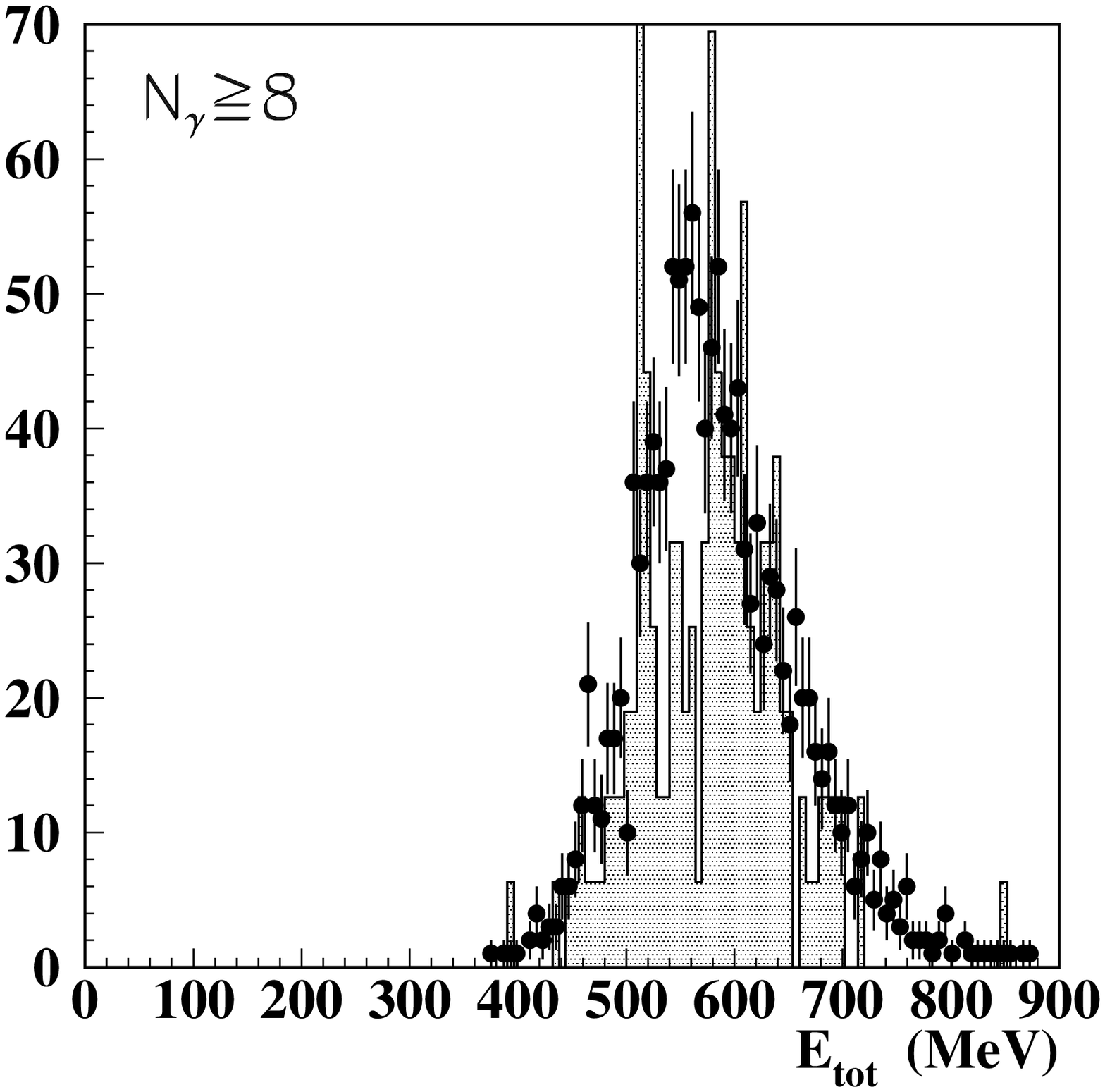,width=6cm}
\end{minipage}
    \caption{$K_L \to \pi^0 \pi^0 \pi^0$ selection:
    distribution of the total energy for events with
    3, 4, 5, 6, 7 and $\geq$ 8 clusters. Dots are data,
    shaded histogram is Monte Carlo simulation for
    $K_L \to$ all channels. Data and Monte Carlo histograms are normalized to
    the same number of entries.}
    \label{FIG2}
\end{figure}

Fig. \ref{FIG2} shows the distribution of the total energy for
events with different numbers of clusters together with the results of the 
Monte Carlo simulation. The relative fraction of events is shown in
Table \ref{TABLE1}. The difference between data and Monte Carlo simulation
for events with $\ge$ 5 clusters is due to split clusters.
The contamination from accidental clusters originated by machine
background is negligible. The residual background contamination in
events with 3 and 4 clusters is evaluated by Monte Carlo simulation and
amounts to $(18.6 \pm 1.0)\%$ and $(7.0 \pm 0.2)\%$ respectively.
\\

\begin{table}[hbt]
\begin{center}
\begin{tabular}{|c|c|c|} \hline
 Number of clusters & Data & Monte Carlo \\ \hline
 3 & $ 0.37 \pm 0.02\ \%$  &  $ 0.35 \pm 0.04\ \% $ \\
 4 & $ 7.2 \pm 0.1\ \%$    &  $ 7.3 \pm  0.1\ \%$  \\
 5 & $ 31.5 \pm 0.1\ \%$   &  $ 32.2 \pm 0.2\ \%$ \\
 6 & $ 57.4 \pm 0.1\ \%$   &  $ 58.4 \pm 0.2\ \%$ \\
 7 & $ 3.3 \pm 0.1\ \%$    &  $ 1.7 \pm 0.1\ \%$ \\
$\geq 8$ & $0.1\ \%$       &  $0.03\ \%$ \\ \hline
\end{tabular}
\end{center}
\caption{ \small Fraction of events with at least three neutral
 clusters connected to the $K_{L}$ decay vertex.}
\label{TABLE1}
\end{table}

A subsample of events has been processed and analysed without
passing through the initial filter. The fractional loss due to the filter is 
found to be less than $10^{-3}$. The trigger efficiency for
$K_{S} \to \pi^{+} \pi^{-}, K_{L} \to \pi^{0} \pi^{0}
\pi^{0}$ events was measured in two different ways. A detailed
description of the methods is given in reference \cite{K-TRIGGER}.
The first method uses only the
data and the information provided by the combined EMC + DC
trigger. In the second method the Monte Carlo is used to evaluate
the correlation between the EMC and the DC trigger showing that
the correlation factor is very small. The results obtained with
the two methods, $\epsilon_{trig1} = (99.88 \pm 0.04)\%$,
$\epsilon_{trig2} = (99.90 \pm 0.03)\%$, are in good agreement.
Since the two methods are independent, and the results consistent, we
combined the two results.

The number of events is:
\[ N_{\pi^{0}\pi^{0}\pi^{0}} = \frac{N_{3}'+ N_{4}' + N_{5} + N_{6} + N_{7}
   + N_{\ge 8} + N_{Dalitz}}
   {\epsilon_{downscale} \cdot \epsilon_{trig} \cdot \epsilon_{sel}}\]
where $N_{3}'$ and $N_{4}'$ are corrected for the background subtraction 
and the additional cuts quoted before. $N_{Dalitz}$, a small addition of
$0.46\%$ of the total count, is obtained using the 
Monte Carlo result that $21.5\%$ of the Dalitz decays are included in
$N_{3} + N_{4}$ events while $60\%$ of them are rejected by the track
veto.
We find
$N_{\pi^{0}\pi^{0}\pi^{0}} = 9,802,200 \times  \ (1 \pm 0.0010_{stat} \pm
0.0016_{syst})$. 

To check the uniformity of the $K_{L} \to
\pi^{0}\pi^{0}\pi^{0}$ vertex reconstruction efficiency throughout the FV
we have studied the proper time distribution.
From a fit to the distribution
we find $\tau_{K_{L}} = 51.6$ ns with a statistical error of 
0.4 ns \cite{K-note}, in good agreement with the value reported in PDG
\cite{PDG2002}, $\tau_{K_{L}} = ( 51.7 \pm 0.4)$ ns.

\section{$K_{L} \to \gamma \gamma$ selection}

$K_{L} \to \gamma \gamma$ events are preselected
by requiring at least
two calorimeter clusters with energy $E_{\gamma} > 100$ MeV not
associated to tracks.  For the two most energetic clusters we
require:
\begin{itemize}
\item[-] total energy, $E_{12} = E_{\gamma1}+E_{\gamma2} > 350 $ MeV;
\item[-] angle between the photon momenta projected onto the plane
normal to the $K_{L}$ direction, $\psi > 150^{\circ}$;
\item[-] time difference smaller than 15 ns;
\item[-] $K_L$ decay vertex in the fiducial volume and $\Delta L_{K} =
|L_{K1} - L_{K2}| < 4 \sigma_{\gamma \gamma}(L_{K})$. 
\end{itemize}

The geometrical acceptance and the selection efficiency are
evaluated by Monte Carlo simulation. The values of the efficiency are shown
in table \ref{TABLE2}.
\begin{table}[h]
\begin{center}
\begin{tabular}{|c|c|} \hline
 preselection & efficiency  \\ \hline
 $E_{\gamma}>100$ MeV)  & (92.6 $\pm$ 0.1$_{stat}$) \% \\
 $E_{12} > $ 350 MeV       & (99.88 $\pm$ 0.02$_{stat}$) \% \\
 $\psi > 150^{\circ}$   & (98.4 $\pm$ 0.1$_{stat}$ $\pm$
 0.4$_{syst}$) \% \\
 $\Delta L_{K} < 4 \sigma $ & (98.5 $\pm$ 0.1$_{stat}$ $\pm$
 0.2$_{syst}$ ) \% \\
 $\Delta t < $ 15 ns & (99.89 $\pm$ 0.02$_{stat}$) \% \\
 \hline
 total   & (89.5 $\pm$ 0.2$_{stat}$ $\pm$ 0.4$_{syst}$) \%
 \\ \hline
 \end{tabular}
 \end{center}
 \caption{\small Efficiency of the $K_{L} \to \gamma \gamma$
 preselection.}
 \label{TABLE2}
\end{table}

With these cuts we obtain 1.7 $\times$ 10$^{5}$ events with a large background due to
$K_{L} \to \pi^{0} \pi^{0} \pi^{0}$ and $K_{L} \to \pi^{0}
\pi^{0}$ decays, $K_{L} \to \gamma \gamma \gamma$ being negligible
\cite{BARR-95-3g}. 

\begin{figure}[htb]
\vfill\begin{minipage}{0.5\linewidth}
  \centering  \epsfig{file=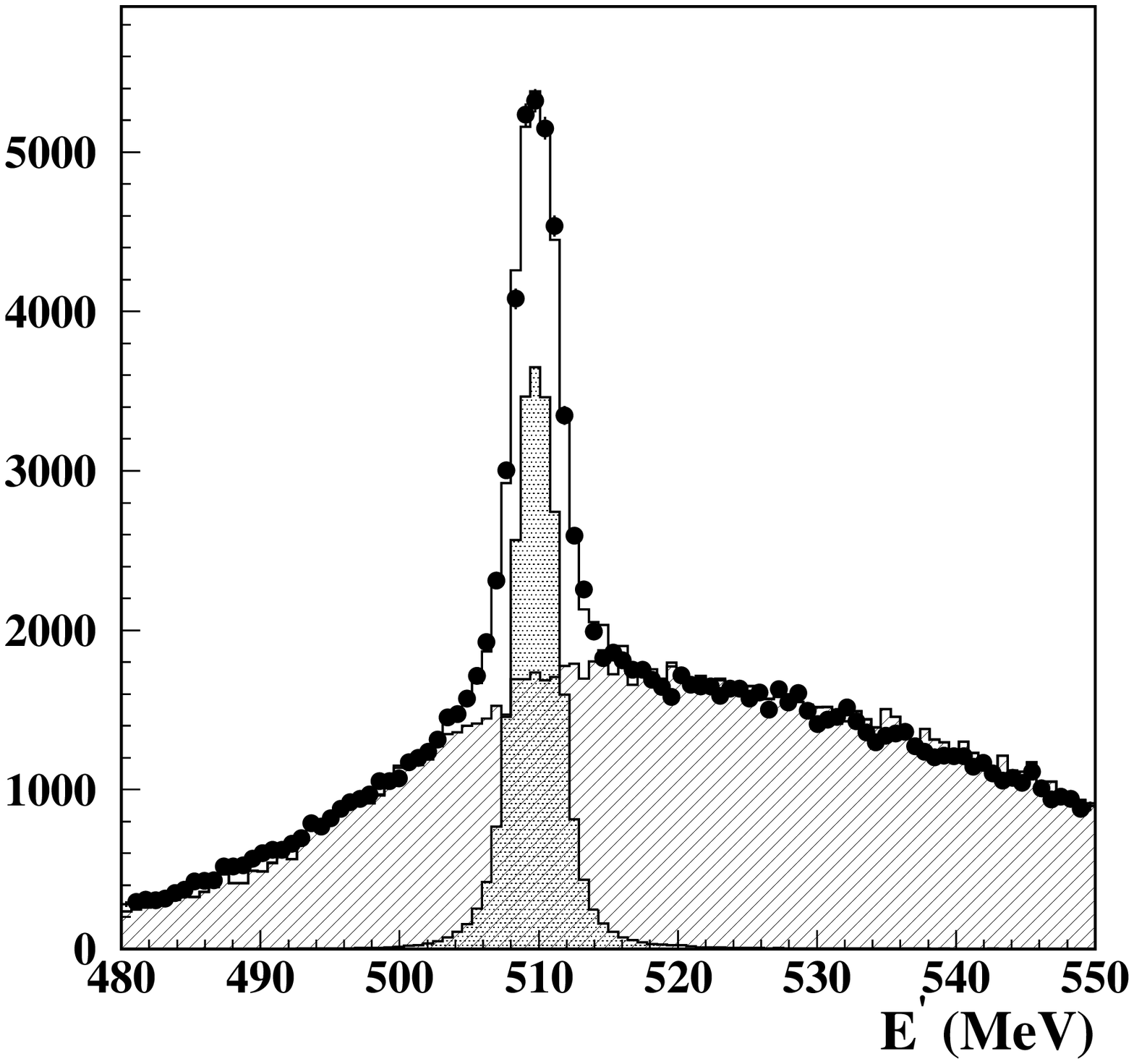,width=7.3cm}
\end{minipage}
\begin{minipage}{0.5\linewidth}
 \centering \epsfig{file=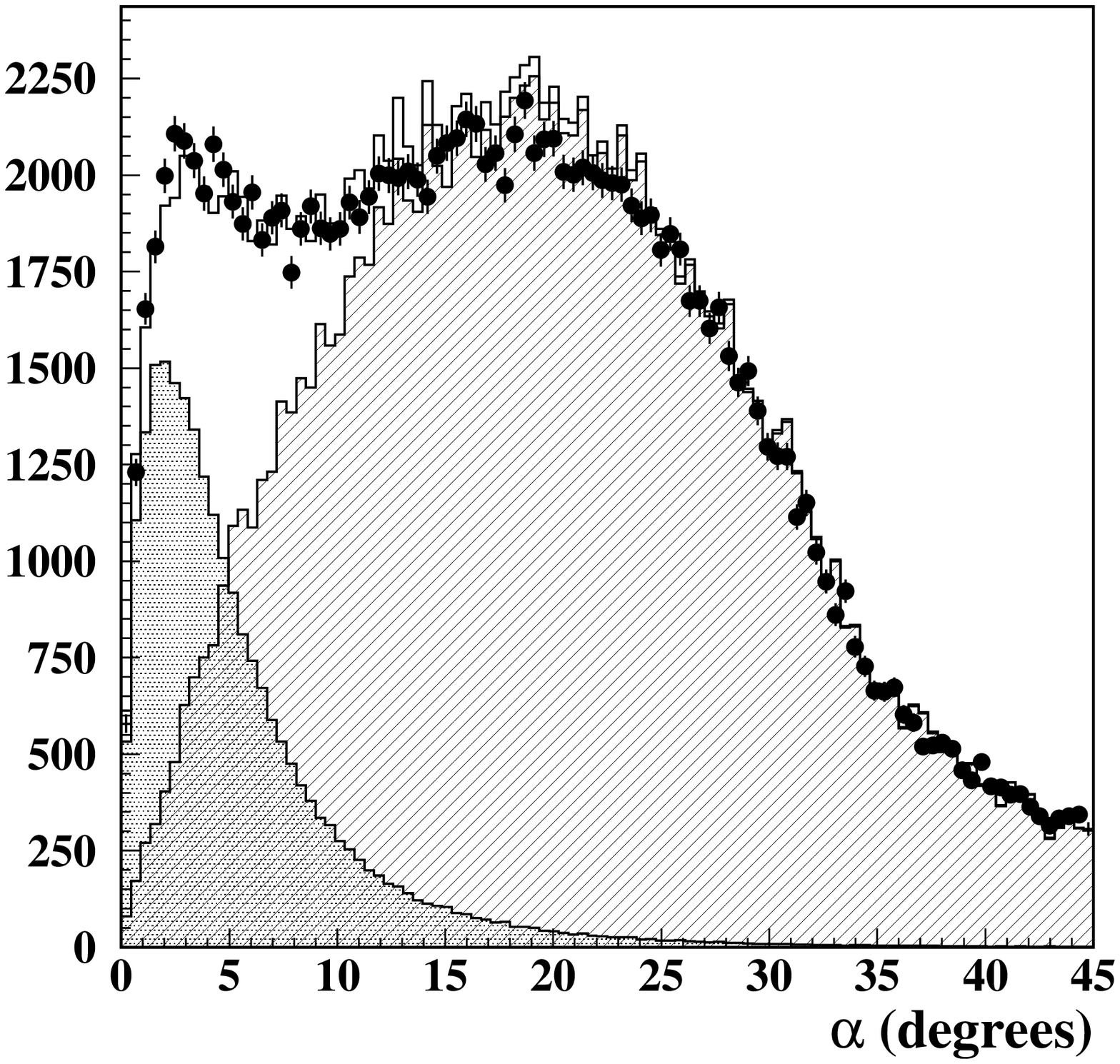,width=7.3cm}
\end{minipage}
 \caption{$K_L \to \gamma \gamma$ selection: 
 distributions of laboratory total energy $E'$ (left) and 
 the angle $\alpha$ between $\vec{p}_{\gamma \gamma}$ and
 $\vec{p}_{K_{L}}$ (right). Dots are data, shaded
 histogram is Monte Carlo simulation for the signal, 
 dashed histogram is Monte Carlo simulation for
 background and solid line histogram is the Monte Carlo simulation 
 for signal and background.}
 \label{FIG3}
\end{figure}

The signal is further selected using the two body
$K_{L} \to \gamma \gamma$ decay kinematics.
In fact, photon energies can be computed with better
accuracy from cluster and decay vertex coordinates. 
The laboratory energy is obtained by boosting 
from the center of mass where 
$E_{\gamma} =  M_{K}/2$ to the laboratory. If $\hat{p}_{\gamma i}$ are unit vectors
from the $K_{L}$ decay vertex to the cluster centroids, the photon
energies are
\[ E^{'}_{\gamma i} = \frac{M_{K} / 2}
   {\gamma_{K} (1 - \beta_{K} \
   \hat{p}_{\gamma i} \cdot \hat{p}_{K})} \]
where $\beta_{K}$ and $\hat{p}_{K}$ are computed from
$\vec{p}_{K_{L}} = \vec{p}_{\phi} - \vec{p}_{K_{S}}$. The $K_{L}$
has energy $E' = E'_{\gamma1} + E'_{\gamma2}$, and
momentum $\vec{p}_{\gamma\gamma} = E'_{\gamma 1} \cdot
\hat{p}_{\gamma1} + E'_{\gamma2} \cdot \hat{p}_{\gamma2}$.
Fig. \ref{FIG3} shows the distribution of $E'$ and of the angle
$\alpha$ between $\vec{p}_{\gamma \gamma}$ and $\vec{p}_{K_{L}}$,
together with the results of the Monte Carlo simulation.
The data are fitted with a linear combination
 of the Monte Carlo distributions for signal and background. The fit
 gives the relative normalisation for the two populations.

In order to reduce background we further require:
\begin{itemize}
\item[-] $| E'  - \mu' | < 5 \sigma'$
where $\mu' = 510.0$ MeV and $\sigma'$ = 1.8 MeV are evaluated
from a fit to the $E'$ distribution;
\item[-] $\alpha < 15^{\circ}$.
\end{itemize}

\begin{figure}[htb]
\vfill\begin{minipage}{0.5\linewidth}
  \centering  \epsfig{file=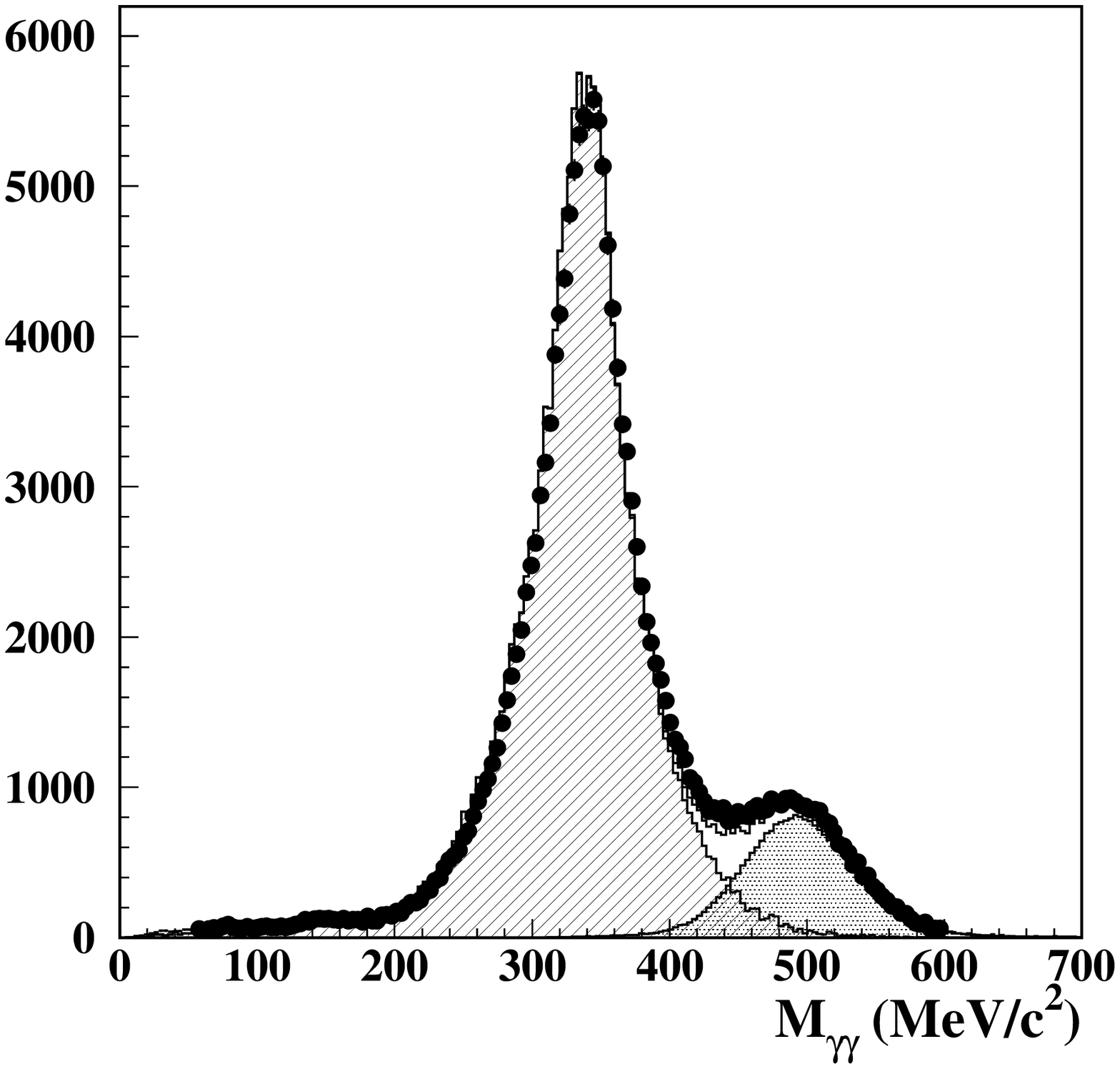,width=7.3cm}
\end{minipage}
\begin{minipage}{0.5\linewidth}
 \centering \epsfig{file=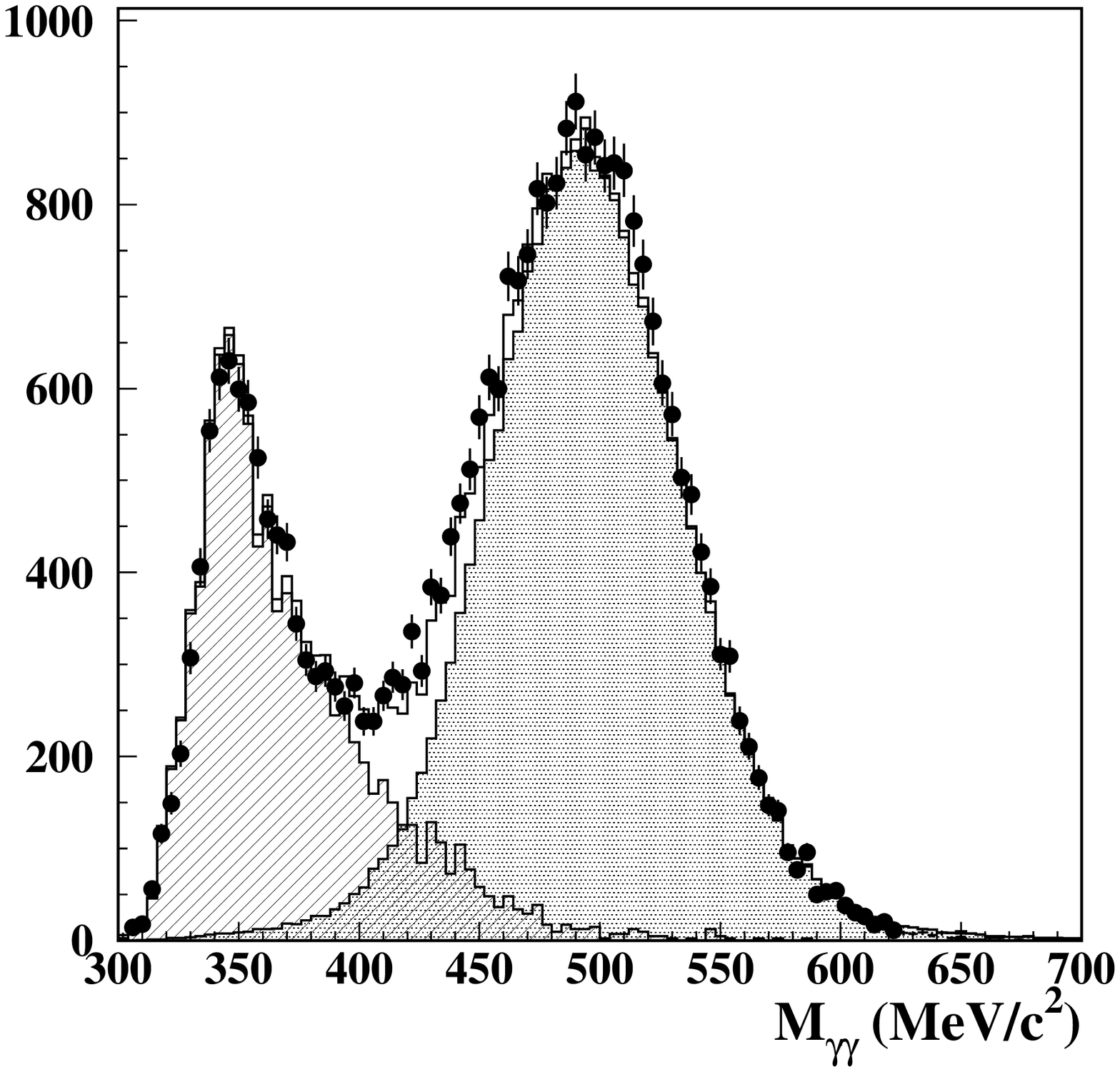,width=7.3cm}
\end{minipage}
 \caption{$K_L \to \gamma \gamma$ selection: 
 distributions of the invariant mass, $M_{\gamma \gamma}$,
 before (left) and after (right) the $E'$ and $\alpha$ cuts.
Dots are data, shaded histogram is Monte Carlo
 simulation for the signal, dashed histogram is Monte Carlo simulation for
 background and solid line is the Monte Carlo simulation for signal
 and background.}
 \label{FIG4}
\end{figure}

To extract the signal we fit the invariant mass $M_{\gamma \gamma}$ 
distribution obtained using calorimeter cluster energies
with a linear combination of the 
Monte Carlo distributions for signal and background. The result of
the fit gives the number of events for the two populations. 
Fig. \ref{FIG4} shows the distribution of $M_{\gamma \gamma}$
before and after the $E'$ and $\alpha$ cuts. 
From a fit to the second distribution
we find $ 22,185 \pm 170$ $K_{L} \to \gamma \gamma$ events.

The efficiency of the selection cuts are evaluated from the data
using a sample of $K_L \to \gamma \gamma$ 
events with high purity (S/B $\sim 10^3$) selected by applying hard,
uncorrelated cuts on other kinematic variables \cite{K-note}.

The systematic error associated with the selection cuts on ($E', \alpha$) is
evaluated by moving the cuts around the chosen values and fitting the
invariant mass distribution.
The maximum displacement of the measured value for the number of signal
counts is $\pm 0.2\%$ and $\pm 0.3\%$ for the $E'$ and the
$\alpha$ distribution respectively. The systematic error due to
the background contamination has been evaluated by changing the shape of
the background distribution used as input of the fit. The effect on the signal is 10
times smaller and produces a systematic error of $\pm 0.3\%$.

The filter and the trigger efficiencies are evaluated as for the
analysis of $K_{L} \to \pi^{0} \pi^{0} \pi^{0}$ decay. The results
are $\epsilon_{filter} = (99.93 \pm 0.01)\%$, $\epsilon_{trig} =
(99.44 \pm 0.04)\%$ where the statistical and systematic errors are
combined in quadrature.
The efficiencies associated with the
various analysis steps are summarised in Table \ref{TABLE3}.
\\

\begin{table}[h]
\begin{center}
\begin{tabular}{|c|c|} \hline
 selection & efficiency \\ \hline
 trigger & (99.44 $\pm$ 0.04) \% \\
 filter & (99.93 $\pm$ 0.01) \% \\
 preselection & (89.5 $\pm$ 0.2$_{stat}$ $\pm$ 0.4$_{syst}$) \% \\
 $|E'-\mu'| < 5 \sigma'$ & (98.5 $\pm$ 0.2$_{stat}$ $\pm$ 0.2$_{syst}$) \% \\
 $\alpha < 15^{\circ}$  &  (92.5 $\pm$ 0.3$_{stat}$ $\pm$ 0.3$_{syst}$) \% \\
 \hline
 total        & (81.0 $\pm$ 0.3$_{stat}$ $\pm$ 0.5$_{syst}$) \% \\
\hline
\end{tabular}
\end{center}
\caption{Efficiencies for the selection of $K_{L} \to \gamma
\gamma$ events. } \label{TABLE3}
\end{table}

The number of events is $N_{\gamma \gamma} = 27,375 \times  \ (1 \pm
0.0076_{stat} \pm 0.0081_{syst})$. For the ratio we find:
\[ R = \frac{\Gamma(K_{L} \to \gamma \gamma)}
   {\Gamma(K_{L} \to \pi^{0} \pi^{0} \pi^{0})} =
   (2.793 \pm 0.022_{stat} \pm 0.024_{syst}) \times 10^{-3} \]
in good agreement with the recent result from the NA48
Collaboration $\Gamma(K_{L} \to \gamma \gamma)/\Gamma(K_{L} \to
\pi^{0} \pi^{0} \pi^{0}) = (2.81 \pm 0.01_{stat} \pm 0.02_{syst})
\times 10^{-3}$ \cite{NA48-2002}. 

Using the known value for the $K_{L} \to \pi^{0} \pi^{0} \pi^{0}$
branching fraction, we
obtain $BR(K_{L} \to \gamma \gamma) = (5.89 \pm 0.07 \pm 0.08) \times
10^{-4}$ 
where the first error represents the 
statistical and systematic error on R combined in quadrature
and the second is due to the 
uncertainty in the  $\pi^0 \pi^0 \pi^0 $ branching fraction.
A decay width of 
$\Gamma(K_{L} \to \gamma \gamma) = (7.5 \pm 0.1) \times 10^{-12}$
eV is in agreement with $\mathcal{O}(p^{6})$ predictions of ChPT
provided the value of the pseudoscalar mixing angle is close to
our recent measurement of $\theta_{P} = (- 12.9^{+
1.9}_{-1.6})^{\circ}$ \cite{ETAMIXING}.

\ack

We thank the DAFNE team for their efforts in maintaining low background 
running  conditions and their collaboration during all data-taking. 

We want to thank our technical staff: 
G.F.Fortugno for his dedicated work to ensure an efficient operation of 
the KLOE Computing Center; 
M.Anelli for his continuous support to the gas system and the safety of the
detector; 
A.Balla, M.Gatta, G.Corradi and G.Papalino for the maintenance of the
electronics;
M.Santoni, G.Paoluzzi and R.Rosellini for the general support to the
detector; 
C.Pinto (Bari), C.Pinto (Lecce), C.Piscitelli and A.Rossi for
their help during shutdown periods.

This work was supported in part by DOE grant DE-FG-02-97ER41027; 
by EURODAPHNE, contract FMRX-CT98-0169; 
by the German Federal Ministry of Education and Research (BMBF) 
contract 06-KA-957;  by Graduiertenkolleg `H.E. Phys. and Part. Astrophys.' 
of Deutsche Forschungsgemeinschaft, Contract No. GK 742; 
by INTAS, contracts 96-624, 99-37; 
and by TARI, contract HPRI-CT-1999-00088.

\end{document}